\documentclass[aps,apl,twocolumn,footinbib,longbibliography]{revtex4-1}
\usepackage{amsmath,amssymb}
\usepackage{bm}
\usepackage{epsfig}
\usepackage{natbib}
\usepackage{hyperref}
\usepackage{color}
\usepackage{graphicx}
\usepackage{gensymb}

\newcommand{\rmi}{{\rm i}}
\newcommand{\comm}[1]{}

\begin{document}

\title{Towards exciton-polaritons in MoS$_2$ nanotubes}

\author{D. R. Kazanov}
\email{kazanovdr@gmail.com}
\author{M. V. Rakhlin}
\author{K. G. Belyaev}
\author{A. V. Poshakinskiy}
\author{T. V. Shubina}
\affiliation{
    Ioffe Institute, 26 Politekhnicheskaya, St Petersburg 194021, Russian Federation
}
\date{\today}

\begin{abstract}
We measure low-temperature micro-photoluminescence spectra along a MoS$_2$ nanotube, which exhibit the peaks of the optical whispering gallery modes below the exciton resonance. The variation of the position and intensity of these peaks is used to quantify the change of the nanotube geometry. The width of the peaks is shown to be determined by the fluctuations of the nanotube wall thickness and propagation of the detected optical modes along the nanotube. We analyse the dependence of the energies of the optical modes on the wave vector along the nanotube axis and demonstrate the potential of the high-quality nanotubes for realization of the strong coupling between exciton and optical modes with the Rabi splitting reaching 400 meV. We show how the formation of exciton-polaritons in such structures will be manifested in the micro-photoluminescence spectra.

\end{abstract}

\pacs{Valid PACS appear here}

\keywords{nanotubes, photonic crystals, TMDC}
\maketitle

\section{Introduction}

Nanotubes (NTs) made of transition metal dichalcogenides (TMD) such as MoS$_2$, WSe$_2$, WS$_2$  (generalized formula is MX$_2$) were first synthesized in the last century \cite{Tenne1992,Remskar1996} and have been intensively investigated since then (see for review \cite{Shubina2019}). The walls of the TMD NTs consist of monolayers connected by a weak van der Waals force \cite{Rao2003}. In the last years, the TMD structures as a whole gained increased attention due to the exceptional optical properties in the monolayer limit. In particular,  the MoS$_2$ monolayer has the direct optical transitions in the visible range that are associated with the A-exciton which have a large oscillator strength~\cite{Arora2015, Robert2018}. Recently, we have shown that the micro-photoluminescence (micro-PL) spectra of the multilayered MoS$_2$ NTs also exhibit the direct exciton emission~\cite{Shubina2019}. Furthermore, the spectra are modulated by pronounced peaks linearly polarized along the tube axis, that were attributed to the optical whispering gallery modes maintained inside the NT wall~\cite{Kazanov2018}.

\begin{figure}[t]
  \includegraphics[width=.99\columnwidth]{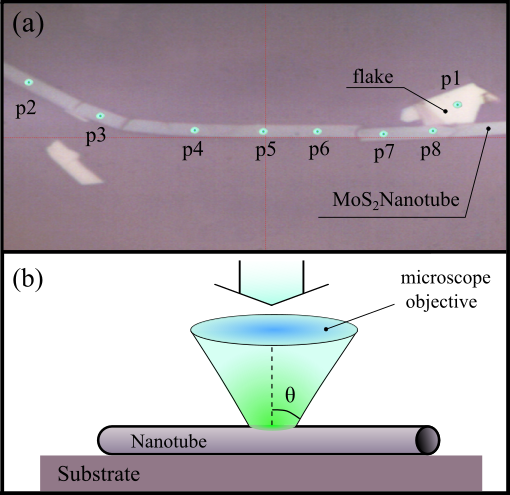}
\caption{ (a) Image of a MoS$_2$ NT taken inside the micro-PL setup. The NT has few cracks that is not essential for our experiment. Green dots indicate spots where the micro-PL was measured. (b) Schematic of NT excitation at micro-PL measurement (not to scale). As shown, the lens collects the emission at angles $\theta \lesssim 25\degree$.
}\label{fig:Scheme}
\end{figure}

The presence of both strong exciton and optical resonances opens a way to couple them when their frequencies coincide. Then, the hybrid polariton modes are formed and the interaction strength can be quantified by the Rabi splitting between them. 
For planar microcavities based on classical semiconductors, the Rabi splitting $\hbar \Omega_{\rm{Rabi}}$ reaches the values of about 10 meV for GaAs, 20 meV for CdTe, 50 meV for GaN, and 200 meV for ZnO~\cite{Wertz2010,Kasprzak2006,Christmann2008,Li2013}. Major interest in exciton-polaritons in 2D van der Waals structures is related to the enhanced interaction between light and matter~\cite{Low2016} as compared to classical bulk materials. In Fabry-Per$\acute{\rm{o}}$t type microresonators with a single MoSe$_2$  monolayer, the Rabi splitting  $\hbar \Omega_{\rm{Rabi}}\sim$ 20 meV was demonstrated. Placing $N$ monolayers inside the cavity enhances the interaction by $\sqrt{N}$, and for four layers of MoSe$_2$ the Rabi splitting of $\sim$ 40 meV \cite{Dufferwiel2015} was achieved. A stronger interaction is realized for microcavities with WS$_2$ monolayers where the Rabi splitting reaches 270 meV for A-exciton and 780 meV for B-exciton \cite{Wang2016} at low temperatures and is about 70 meV \cite{PLatten2016} at room temperature. 
Geometry more complex than the planar allows for fine tuning of polariton spectra. 
The interaction of excitons with the waveguide mode in the structures based on MoSe$_2$ was demonstrated to produce the Rabi splitting of $\sim $ 100 meV \cite{Hu2017}. Possible Rabi splitting of about 280 meV was derived via analysis of extinction spectra of an ensemble of WS$_2$ nanotubes \cite{Yadgarov2018}.

\begin{figure*}[t]
\centering
  \includegraphics[width=1.99\columnwidth]{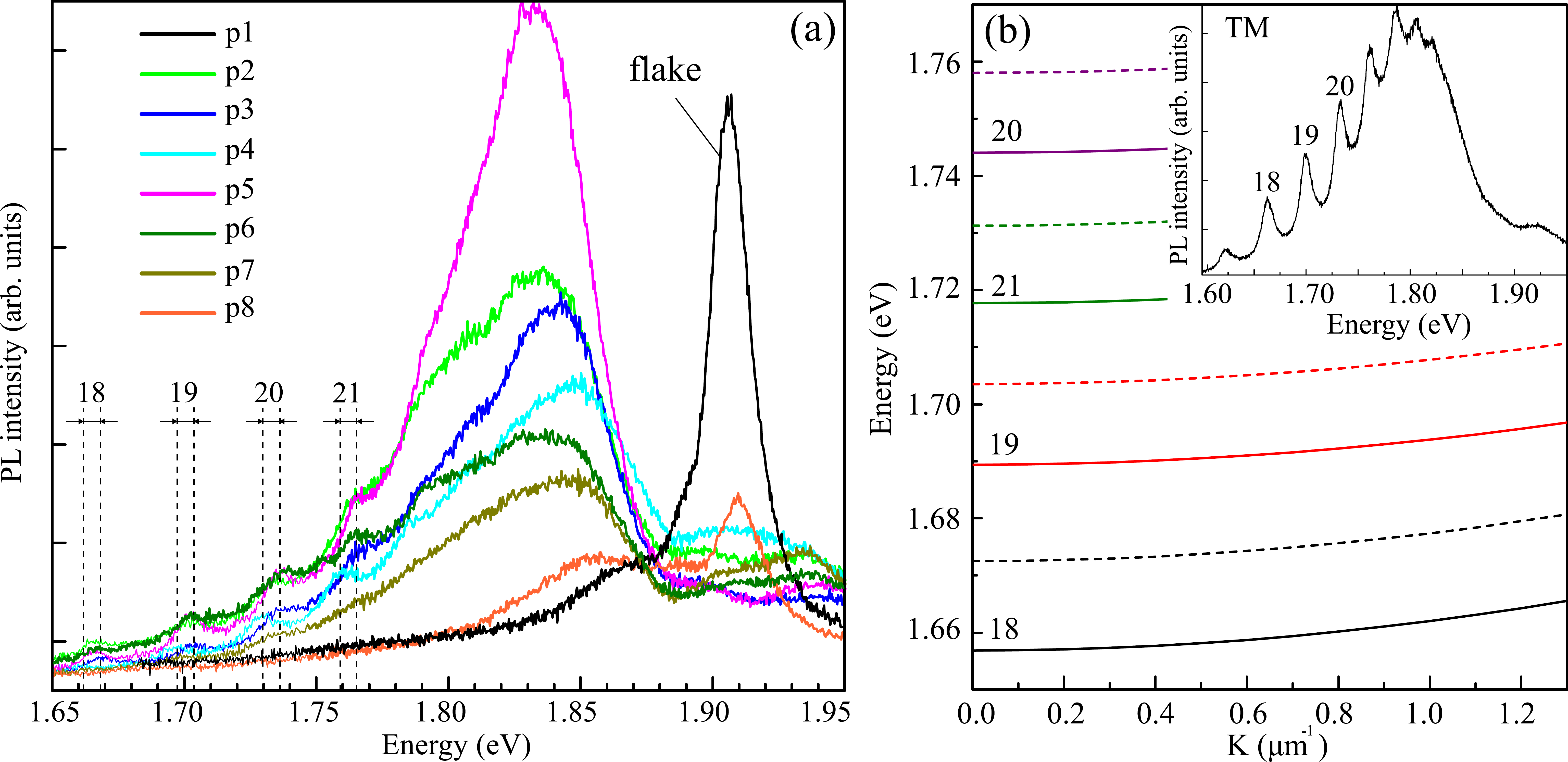}
\caption{ (a) Non-polarized micro-PL spectra measured at 10~K in the spectral range of direct exciton transitions from several spots of MoS$_2$ NT, indicated by green dots in Fig.~\ref{fig:Scheme}a. Black line is a spectrum taken from the planar MoS$_2$ layer (p1, flake). The dotted lines indicate the energies of the TM-polarized whispering gallery modes with angular quantum numbers $m=18-21$ calculated for the NT with the wall thickness varying by 1 monolayer. (b) Dependence of the energies of the optical modes with the angular numbers $m=18-21$ on the wave vector $K$ along the NT axis calculated for NTs with the 44-monolayers wall (dashed lines) and 46-monolayers wall (solid lines). The inset shows the TM-polarized micro-PL spectrum of the NT at 80 K when the intensity of basic exciton peak quenches. }
\label{fig:PLSpectra}
\end{figure*}

In this paper, we investigate micro-PL of MoS$_2$ NTs. We model the micro-PL spectra and reproduce the peaks associated with whispering gallery modes. Our studies reveal that the position of peaks fluctuates along the NT due to the variation of the wall width, whereas the broadening of the peaks is determined by the dispersion of the optical modes, whose impact occurs via the finite numerical aperture of used objective. We propose to use NTs of high quality as a microresonator to realize the strong coupling between the optical modes and excitons and show how formation of the exciton-polaritons would transform the PL spectra.

\section{Experiment and modeling of optical modes}

To study the optical properties of single MoS$_2$ NTs, a micro-PL experiment was performed. A sample with NTs on a Si substrate~\cite{Kazanov2018} was mounted in a He-flow cryostat with an Attocube XYZ piezo-driver inside, which allowed to precisely maintain the positioning of a chosen part on a NT with respect to a laser spot. Micro-PL measurements were performed at low temperature of 10~K, when the direct exciton emission prevails~\cite{Shubina2019}. Focusing of a laser beam on the sample was carried out using a 50-fold objective (Mitutoyo 50xNIR, $\rm{NA}=0.42$), which also was used to collect the PL signal. The enlarged image of the sample was transferred by means of achromatic lens to the plane of a mirror with a calibrated aperture (pinhole - 200 $\mu$m). This arrangement  determined the region on the sample from which the micro-PL signal is detected. To record the PL spectra from the NT the collected signal passed through the gratings of the monochromator and entered the CCD camera. A 405-nm line of a semiconductor laser was used for non-polarized PL excitation. The laser power density correspond to $\sim$10 mW per area of about 20---50 $\mu$m$^2$.

Figure~\ref{fig:Scheme}a shows a photo of a NT taken inside of the micro-PL setup. The NT is about 50 microns in length and 2 microns in diameter. The green dots p1---p8 show the locations from which the micro-PL signal was recorded. Their size corresponds approximately to the size of the signal detection spot in the experiment. The number of layers in the NT wall has been determined by modeling the micro-PL spectra, as described in Ref.~\cite{Kazanov2018}, which yields the average value of about 45 monolayers.

The micro-PL spectra in the optical range of the direct exciton transitions measured in different spots of the NT are shown in Fig.~\ref{fig:PLSpectra}a. The black line indicates the spectrum taken from the surface of the plane layer (flake) featuring a bright inhomogeneously broadened exciton resonant peak. The major peaks in the micro-PL spectra recorded from NT are wider and red shifted by $\sim 50$ meV  with respect to the flake due to  uncompensated stresses and 3R-polytype stacking of the monolayers in the wall of the chiral NT~\cite{Wilson1969,Shubina2019}. While the peaks at about $\sim$~1.83~eV correspond to A-exciton in MoS$_2$, the series of smaller peaks at lower energies stems from TM-polarized optical whispering gallery modes in the NT. Here, they are characterized by angular quantum number $m=18 $---$ 21$. Modes with energies higher than the exciton energy are not observed due to the strong absorption. TE-polarized whispering gallery modes are absent in the spectrum due to their lower $Q$-factor that makes them much less pronounced. 
The black dashed lines in Fig.~\ref{fig:PLSpectra}a indicate the maximum deviation in the positions of the peaks detected from different spot of the NT. For all of the observed modes, the spread is around 6 meV, which corresponds to a change of the wall thickness by 1 monolayer. 

The intensity of the PL is maximal in the center of the NT, point p5 in Fig.~\ref{fig:Scheme}a, and decreases at the NT end. Analyzing the evolution of the spectra from point p5 to point p8 in Fig.~\ref{fig:PLSpectra}a, we can assume that the geometry of the NT is adiabatically changing as it becomes less like a NT and more like a ribbon (flattened tube). This is also confirmed by a decrease in the intensity of the peaks associated with the presence of optical modes and an increase in the intensity of the peak associated with the flat layer. It should be noted that NTs start to form inside microfolds or bend edges of curved flakes~\cite{Remskar1996}. To put a NT on the SiO$_2$ substrate one should tear it from the the silica ampoules where the NTs were grown. The point p8 corresponds to the tip of the NT, which consists of a chunk of the flake and the incipient part of the NT. Thus, due to the limited spatial resolution, we observe PL contributions from both the NT and the flake which has approximately the same thickness.
The other end of the tube (points p2--p4) turns out to be undamaged, as we observe pronounced optical modes in the corresponding spectra. In addition, we have shown TM-polarized micro-PL spectrum where the peaks corresponding to whispering gallery modes are more pronounced (see the inset in Fig.~\ref{fig:PLSpectra}b). 

To model the PL spectra, we solve Maxwell's equations for the hollow cylinder with a certain inner and outer radii, dielectric permittivity of MoS$_2$, and the homogeneously distributed sources of radiation. Generalizing the approach developed in~\cite{Kazanov2018}, we  calculate the radiation intensity not only in the direction perpendicular to the NT axis, but also in other directions, characterized by angle $\theta$, see Fig.~\ref{fig:Scheme}b. In the experiment, the PL from the directions with $\theta \lesssim 25^\circ$ is collected by the microscope objective. Therefore, the PL spectra is contributed by the optical modes of frequency $\omega$ corresponding wave vector along the tube axis $K=(\omega/c) \sin \theta \lesssim 3.5 \mu$m$^{-1}$ ($c$ is the speed of light). The spread of their energies widens the PL peaks. 

Figure \ref{fig:PLSpectra}b shows the dependence of the energies of the optical modes with angular numbers $m=18-21$ on the wave vector along the NT axis $K$. Calculation was made for NTs with the wall consisting of 44 (dashed lines) and 46 monolayers (solid lines). Reduction of the NT wall thickness leads to the increase of the mode energies due to the stronger confinement. The dependence of the mode energies on the wave vector $ K$ can be expressed as $ E(K) = \hbar c/\textit{n}_{\rm{eff}} \sqrt{K^2 +(m/R)^2}$, where $R$ is the NT radius, $n_{\rm{eff}} $ is the effective refractive index, which can be estimated as $n_{\rm{eff}} \approx n\eta$ with $\eta \approx 0.44$ being the part of the electric field that is confined inside the NT wall.
For small wave vectors $K \ll m/R$ considered here, the dispersion is quadratic, $E(K)=E(0) + \hbar^2 K^2/2M^*$, with the effective mass $M^* = \hbar m n_{\rm{eff}} / (R c) \approx 7.7 \cdot 10^{-6} m_0$, where $m_0$ is the mass of the free electron. Numerically calculated dispersion shown in Fig.~\ref{fig:PLSpectra}d yields the close value $M^* = 7.4\cdot 10^{-6} m_0$. 
Then, the PL peak broadening due to finite spread of detection angles is estimated as $\delta E \approx E^2\sin^2 \theta /(2M^*c^2) = 55\,$meV. This value shows the upper bound of the peak width observed in the experiment. 
To sum up all of the above, PL spectra indicate that the actual tube of MoS$_2$ is not homogeneous. The observed variation of the peak positions in the PL spectra from different parts of the NT is due to the fluctuations of the NT wall thickness. The PL peak width is additionally contributed by the finite PL detection angle. 

\begin{figure*}[t] 
\centering
\includegraphics[width=1.99\columnwidth]{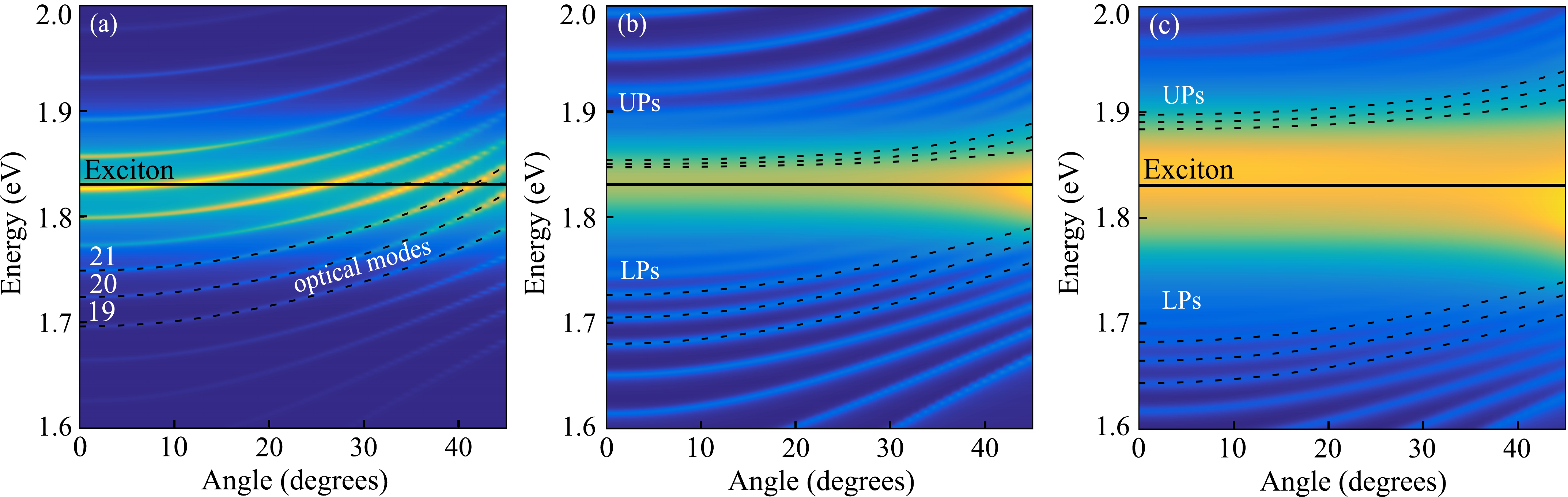}
\caption{Color plot of the PL spectra dependence on the detection angle calculated for the NTs without inhomogeneous broadening of the exciton resonance and different values of Rabi splitting, 2 meV, 20 meV,  and 200 meV (from left to right). The dotted lines show the dispersion of the eigenmodes with the angular  quantum numbers $m = 19,~20,~21$. The black solid line indicates the energy of the exciton. UPs and LPs denote upper and lower polariton modes, respectively.}
\label{fig:Polaritons}
\end{figure*}

\section{Polaritons in nanotube resonator}

In the above considerations, the exciton resonance was assumed to have strong inhomogeneous broadening. Qualitatively new effects are expected  in the higher-quality structures with sharp exciton resonances where the strong coupling regime between the optical modes of the NT and the excitons can be realized. The important role is played by the wave vector along the NT axis $K$ that allows for the fine tuning of the optical mode energy to match the exciton energy. Here, we discuss the properties of such hybrid exciton-polariton modes in such case and show how they would be manifested in the PL spectra.

 To calculate the dispersion of exciton-polaritons in the NT resonator, we assume that its walls are characterized by the local single-pole dielectric function
\begin{equation}
\varepsilon(\omega) = \varepsilon_{b}\left(1 + \frac{\omega_{\rm{LT}}}{\omega_{\rm{ext}}-\omega-\rmi \Gamma}\right) \,,
\end{equation}
where $\omega_{\rm{LT}}$ is a longitudinal-transverse splitting, $\omega_{\rm{ext}}$ is the exciton peak frequency. We estimate its value from the known radiative exciton life time  of $1/(2\Gamma_0) = 0.23$\,ps  in the MoS$_2$ monolayers~\cite{Palummo2015}. Assuming that the NT wall consists of dozens of isolated monolayers (the certain isolation can be supposed for the strained chiral tubes), the   longitudinal-transverse splitting is calculated as~\cite{Ivchenko2005}
\begin{equation}
\omega_{\rm{LT}} = \frac{2 \Gamma_0 c}{\omega_{\rm{exc}} d \sqrt{\varepsilon_{b}}} \,,
\end{equation}
where $\varepsilon_{b} = 16.2 $ is the background dielectric constant of MoS$_2$ and $d = 6.7$\,\AA\ is the interlayer distance, yielding $\hbar \omega_{\rm{LT}} \sim 114$ meV. The strength of the interaction of optical mode and exciton can be quantified by the value of the Rabi splitting between the energies of the hybrid polariton modes, which are formed when the energies of bare excitations coincide. 

The interaction value under the conditions of the exact resonance of the exciton energy $E_{\rm{exc}}(K)$ and the optical mode mode $E_m(K)$ is denoted by the Rabi splitting $\hbar \Omega_{\rm{Rabi}}$. Taking into account that only the fraction $\eta\approx 0.44$ of the optical mode has the electric field inside the NT wall and  can interact with excitons~(see Ref.~\onlinecite{Kaliteevski2007} for a similar consideration for polaritons in cylindrical cavities), the Rabi frequency is calculated as
\begin{equation}
\Omega_{\rm{Rabi}} = \sqrt{2 \omega_{\rm{ext}} \omega_{\rm{LT}}\eta} \,,
\end{equation}
which yields $\hbar \Omega_{\rm{Rabi}}\sim$ 400 meV. The Rabi splitting in the considered structure surpasses the typical values for the resonator structures based on classical semiconductors such as GaAs or GaN. This highlights the potential of the high-quality NT resonators based on MoS$_2$ for realization of strong light-exciton interaction. In the state-of-the-art structures the potentially large Rabi splitting turns out to be masked by even stronger inhomogeneous broadening. 

To show the effect of strong light-exciton interaction, we calculate the PL spectra as a function of detection angle for different value of the Rabi splitting, see Fig.~\ref{fig:Polaritons}. In the regime of weak interaction, see panel~(a) corresponding to $\hbar\Omega_{\rm Rabi} =2$\,meV, the PL spectra consists of the peaks corresponding to the optical modes with different angular numbers $m$. The intensity of the peaks grows when they approach exciton resonance energy, but their energies (dashed lines) remain unperturbed.  
The increase of the light-exciton interaction strength leads to appearance of a series of anti-crossings, each of them reflecting the interaction of an optical mode and the excitonic mode with the same angular quantum number $m$, see panels~(b) and~(c) corresponding to $\hbar\Omega_{\rm Rabi} =20$ and 200\,meV. The modes with different $m$, which were degenerate in the absence of interaction, are now split into upper  and lower polariton modes and a broad peak at the exciton frequency appears between them. 
%

\section{Summary}

In this work, we have studied the micro-PL from the MoS$_2$ NTs. The spectra of micro-PL from different parts of the NT, as well as from the planar layer MoS$_2$ (flake) were compared. In addition to the main peak associated with the optical A-exciton transition, the peaks corresponding to the whispering gallery modes, dominating in TM polarization,  were observed. Variation of the PL spectra measured from different parts of the NT indicates the change of the NT geometry from the cylindrical tube to collapsed ribbon-like one.  We also observed a shift of the energies of the optical modes due to the variation of the number of monolayer in the NT wall. The peaks are additionally broadened due to the dependence of the optical mode energies on the wave vector along the NT axis and the finite spread of the PL detection angles. We have predicted the high potential of high-quality NT with low inhomogeneous broadening for realization strong coupling between light and excitons. We have calculated the PL spectra of such structures and discussed the signatures that will allow for experimental identification of exciton-polaritons in NTs, once the quality of the structures is improved. 

\section*{Acknowledgement}

The work was supported by the Russian Science Foundation (project $\#$ 19-12-00273). A.~V.~P. acknowledges the partial support from RFBR grant No.~18-32-00486, the Russian President Grant No. MK-599.2019.2, and the Foundation ``BASIS''. The authors thank M.~Rem\v{s}kar and S.~Fathipour for providing the samples.

\end{document}